\documentstyle[12pt]{article}
\newcommand{\beq}{\begin{equation}}
\newcommand{\eeq}{\end{equation}}
\newcommand{\bea}{\begin{eqnarray}}
\newcommand{\eea}{\end{eqnarray}}
\begin{document}
\thispagestyle{empty}
\vspace*{-15mm}
\baselineskip 10pt
\begin{flushright}
\begin{tabular}{l}
{\bf OCHA-PP-186}\\
{\bf February 2002}\\
{\bf hep-th/0202050}
\end{tabular}
\end{flushright}
\baselineskip 24pt 
\vglue 10mm 
\begin{center}
{\LARGE\bf{Exciton in Matrix Formulation\\
of Quantum Hall Effect
}}
\vspace{7mm}\\
\baselineskip 18pt 
{\bf
Yuko KOBASHI, Bhabani Prasad MANDAL\footnote{JSPS Postdoctoral Fellow}, and Akio SUGAMOTO}
\vspace{2mm}
{\it 
 Department of Physics, Ochanomizu University, \\
 2-1-1, Otsuka, Bunkyo-ku, Tokyo 112-8610, Japan
}\\
\vspace{10mm}
\end{center}
\begin{center}
{\bf Abstract}\\[7mm]
\begin{minipage}{14cm}
\baselineskip 16pt
\noindent
The quantum Hall effect is studied by introducing two different matrix variables for electrons and holes, having Chern-Simons type interactions.  By generalizing the  constraint condition proposed by Susskind to realize the Pauli's exclusion principle in this two component matrix model, the classical exciton solution having excitation of both quasi-electron and quasi-hole is obtained.  The constraint condition is also solved quantum mechanically in the  infinite-sized matrix case, giving the examples of the physical states. Using these quantum states, the corrections of the exciton energy, coming from the noncommutativity of space (Pauli principle) and from the quantum effects of the background state, are estimated in the lowest order perturbation expansion.  As a result, the dispersion relation of exciton is obtained.
\end{minipage}
\end{center}
\newpage
\baselineskip 18pt 
\def\thefootnote{\fnsymbol{footnote}}
\setcounter{footnote}{0}
\section{Introduction}
In a beautiful paper, Susskind~\cite{Susskind} has proposed a matrix model of the Chern-Simons type in the study of Quantum Hall Effect.  In his approach, the coordinates $\{x^{a}(t)\}~(a=1, 2)$ of the two dimensional multi-electron system are represented by the infinite-sized hermitian matrices $\hat{x}^{a}(t)~(a=1, 2)$.  The matrices evolve in time under the influence of the Lorentz force given by the strong magnetic field $B$ which is supplied from outside in the normal direction to the two dimensional sample of the quantum Hall effect.  Due to the Pauli's exclusion principle it is well known that each electron doing the cyclotron motion cannot overlaps with each other.  Susskind has introduced this exclusion principle by the following constraint condition in the matrix form:
\beq
[ \hat{x}^a,  \hat{x}^b]= i\theta\epsilon^{ab}\hat{\bf{1}},
  \label{constraint}
\eeq
where $\epsilon^{ab}$ denotes the usual antisymmetric tensor, and $2\pi \theta$ is the mean area to be occupied by an electron, namely, the inverse of the number density $\rho$ of the electron,
\beq
2\pi \theta= \rho^{-1}.
\label{2}
\eeq
An analogy in quantum mechanics tells that $\theta$ plays the role of $\hbar$, so that the minimum area to be occupied by an electron in the "phase space" $(x^1, x^2)$ becomes naturally $2\pi \theta$ as is required.

The quasi-electron or the quasi-hole in the quantum Hall system is given, respectively, by the surplus or the deficit of area $q$ occupied by the electrons~\cite{Susskind}.  If this area is fixed by the quantization condition of the magnetic flux passing through the area, we have 
\beq
eBq= 2\pi \times n~~ (n:\mbox{integer}),
\label{flux quantization}
\eeq
so that the charge of the quasi-electron or the quasi-hole is quantized in the unit of $\nu e$:
\beq
e_{qe~or~qh}= \nu e \times n~~ (n: \mbox{negative or positive integer}),
\eeq
where $\nu$ is the filling fraction.  It measures how many electrons exist in the area through which the unit quantum of magnetic flux penetrates, that is,  
\beq
\nu=\rho \frac{2\pi}{eB}.
\label{sec1.5}
\eeq
Now the surplus or deficit area reads from Eq. (\ref{flux quantization}) as
\beq
q=2\pi \theta \nu \times n.
\label{sec1.6}
\eeq

This Susskind approach is reformulated using finite-sized matrix with an additional boundary field in \cite{Polychronakos}, and it is pointed out there that the finite matrix model becomes identical to the solvable Calogero model.  This finite matrix model for the quantum Hall effect is further developed by introducing the coherent states, using which the relation between the physical states of the matrix model and the Laughlin's wave functions is studied~\cite{Sakita}. 

The solution of the constraint equation for the quasi-electron or the quasi-hole is already known in \cite{Susskind}, but the exciton solution of having both quasi-electron and quasi-hole at the same time is not known.  We may recognize that in order to find the exciton solution in the matrix model, we have to give up the hermiticity of the operators and introduce the "unphysical" negative probability.  This reminds the introduction of positron (the hole or the anti-particle of electron) to solve the negative energy difficulty in QED.  There, the annihilation of the "unphysical" electron with negative energy is replaced by the creation of the physical positron (hole) with positive energy.  Namely, there is a possibility in our problem that the surplus (quasi-electron) and the deficit (quasi-hole) of area are realized by the electron and the hole fields separately.  
( See the details in Appendix).

With this motivation, we introduce in this paper two kinds of matrices, the matrices for electrons $\hat{x}^a_e(t)$, and those for the holes $\hat{x}^a_h(t)$ from the beginning, and study the exciton solution and its dispersion relation.  We consider here the infinite matrix case of Susskind without introducing the boundary field as in \cite{Polychronakos}. 
Introduction of both electron and hole fields gives two sets of creation and annihilation operators $(a_e, a^{\dagger}_e)$, and $(a_h, a^{\dagger}_h)$.  This is also beneficial to understand the Laughlin wave functions \cite{Laughlin1}.  Having these two sets of creation and annihilation operators,
the Hamiltonian $H$ and the 3rd component of the angular momentum of the quantum Hall effect $M$ are given by 
\bea
H&=& \sum_i \omega_e (a^{\dagger}_e a_e +1/2)_i + \sum_i \omega_h (a^{\dagger}_h a_h+1/2)_i + \mbox{Coulomb potentials}, \\
M&=&\sum_i (a^{\dagger}_e a_e)_i - \sum_i (a^{\dagger}_h a_h)_i,
\eea
where the cyclotron frequencies are given by
\beq
\omega_e= eB/m_e, ~~\mbox{and}~~ \omega_h= eB/m_h,
\eeq
with the effective masses $m_e$ and $m_h$ for electron and hole, respectively. In this paper, we always set $\hbar=1$ and $c=1$.

If the Coulomb potentials can be ignored, the single particle eigenstates are given by 
\beq
| m, n >\propto (a^{\dagger}_h)^{m}(a^{\dagger}_e)^{n}| 0; 0 >.
\eeq
Then, for the N electron system, the Laughlin wave function $\psi_m$ of the filling fraction $\nu=1/m$ is given by a certain antisymmetrization in the product of the single particle wave functions:
\bea
  \psi_m &\propto&  \prod_{i \le j}^N \left( a^{\dagger}_{h i }- a^{\dagger}_{h j }\right)^m | 0, \dots, 0;  0, \dots, 0> 
\label{sec1.11}
\\
&\propto&
\displaystyle{\sum_{\{i\}}^N}\left( \epsilon^{i_1, i_2, \dots, i_N} (a^{\dagger}_{hi_1})^0  (a^{\dagger}_{hi_2})^1 (a^{\dagger}_{hi_3})^2 \cdots  (a^{\dagger}_{hi_N})^{N-1}  \right)^m | 0, \dots, 0;  0, \dots, 0>,\nonumber\\
\label{sec1.12}
\eea
where the second expression is by the van der Monde determinant.  The important point to be stressed here is the following: To obtain the less filling fraction states without increasing the Landau levels, we have to excite not the electron quanta but the "hole quanta".  Since if $\omega_e \gg \omega_h$ holds or if we do not care about the hole energy, the excitation of the holes do not change the energy, but increases the absolute value of the angular momentum.  Electron with higher angular momentum is moving on a circle with the $|M|$ times larger radius, reducing the filling number, or in other words, the $m$ excitations of the holes cancel the minimum sized $m$ cyclotron motions of electrons, remaining the large sized cyclotron motion with lower filling number.

In the next section, we introduce the two component matrix model and obtain the exciton solution classically.  Quantization is examined using two kinds of descriptions called "1d" and "2d" pictures.  In Sec. 3, physical states of the model are studied at the quantum level.  The energy of the exciton and its dispersion relation is studied in Sec. 4.  The last section is devoted to discussion.
\section{Two Component Matrix Model of Quantum Hall Effect and its Exciton Solution
}
As was stated in the introduction, we start with two component matrix model of the Chern-Simons type, by introducing two independent matrices for electrons and holes. It gives simply the motion under the Lorentz force caused by external magnetic field. Then the starting Lagrangian is written as follows:
\begin{eqnarray}
 L_0={\int}dt\frac{eB}{2}\epsilon_{ab}  [Tr\{(\dot{\hat{x}}^a_e+i[\hat{x}^a_e,\hat{A_0}])\hat{x}^b_e \}
            \nonumber\\
        -Tr\{(\dot{\hat{x}^a_h}+i[\hat{x}^a_h,\hat{A_0}])\hat{x}^b_h \}
-2\theta Tr\{\hat{A_0}\}].
  \label{e12}
\end{eqnarray}
Hereafter, we will specify the infinite sized matrix variables by putting the hat on them.

In the above Lagrangian, an auxiliary gauge potential $\hat{A_0}$ is introduced so as to give the reasonable constraint coming from the Pauli's exclusion principle.
From the starting Lagrangian the constraint condition is obtained as the equation of motion for $\hat{A}_0$:
\beq
[\hat{x}^1_e, \hat{x}^2_e]-[\hat{x}^1_h, \hat{x}^2_h]=i \theta\hat{\bf{1}}.
\label{constraint of the 2 matrix mdel}
\eeq
This constraint means that the area occupied by electrons contributes positively to the constraint, while that by holes does negatively, so that the minimum value of the difference of two (phase space) areas is to be $2\pi \theta$.


To see the meaning of the constraint condition more clearly, it is helpful
to consider the limit of $\theta \to 0$, in a similar way to the classical
limit of $\hbar \to 0$ in quantum mechanics.  Then, the infinite matrices,
$\hat{x}_{e}$ and $\hat{x}_{h}$ become the ("classical") functions of
"phase space" variables $(y^1, y^2)$ and $t$, namely, $x_{e}=x_{e}(t, y^1,
y^2)$ and $x_{h}=x_{h}(t, y^1, y^2)$.  This is the continuous fluid
description of the quantum hall effect, where the position of the electron
fluid $x^a_{e}$ and that of the hole fluid $x^a_{h}$ are described by the
original position at $t=0$, namely, $(y^1, y^2)$.
 
At the same time, the commutator becomes $i\theta$ times the Poisson
bracket(Jacobian of the "phase space" variables), and the $Tr$ becomes the
"phase space" integral.  Therefore, in this continuous limit, the
constraint becomes
 
$$ \frac{\partial(x^1_{e}, x^2_{e})}{\partial(y^1,
y^2)}-\frac{\partial(x^1_{h}, x^2_{h})}{\partial(y^1, y^2)}=1$$.
 
This shows that the volume occupied by the electron fluid minus that by the
hole fluid is always equal to the original volume labeled by $(y^1, y^2)$,
showing the incompressible fluid.  As is seen from the expression, the
constraint condition permits many solutions having the equal difference,
but the hole fluid represents the deficit of the electron fluid, so that
the difference gives the density of the electron fluid itself which is kept
constant except at the position of quasi-electron or quasi-hole.


If the fractional surplus of area $q=2\pi\nu$ discussed in Eq.(\ref{sec1.6}) occurs at the quasi-particle's excitation position and the deficit of the same area occurs at the quasi-hole's excitation position, we  have the classical exciton solution.  As is mentioned in the Introduction
and the supplement given in Appendix, to express the solution, we introduce two kinds of lowering operators in the matrix space, $\hat{b}$ and $\hat{d}$.  If the infinite dimensional vector space on which all the matrices operate is spanned by $\{ |0>, |1>, \dots \}$, then  $\hat{b}$ and $\hat{d}$ are defined as follows:
\bea 
\hat{b}^{\dagger}|n>&=&\sqrt{n+1+\nu}|n+1>   \nonumber \\
\hat{b}|n>&=&\sqrt{n+\nu}|n-1> ~~\mbox{for}~~ n \ne 0, \nonumber \\
\hat{b}|0>&=&0,
\label{sec2.15}
\eea
and 
\bea 
\hat{d}^{\dagger}|n>&=&\sqrt{n+1-\nu}|n+1>   \nonumber \\
\hat{d}|n>&=&\sqrt{n-\nu}|n-1> ~~\mbox{for}~~ n \ne 0, \nonumber \\
\hat{d}|0>&=&0.
\label{sec2.16}
\eea
Then, the classical exciton solution $(\hat{z}_e, \hat{z}_h)_{cl}$ in the complex notation is obtained:
\bea
(\hat{z}_e)_{cl} &\equiv& (\hat{x}^{1}_{e} + i \hat{x}^{2}_{e})_{cl} = z_e(t) \hat{1}+ \sqrt{2\theta} \hat{b}, \nonumber \\
&=& z_e(t) \sum_{n=0}^{\infty}|n><n|+ \sqrt{2\theta}\sum_{n=1}^{\infty} \sqrt{n+\nu}|n-1><n|,  
 \label{sec2.17}
\\
(\hat{z}_h)_{cl} &\equiv& (\hat{x}^{1}_{h} + i \hat{x}^{2}_{h})_{cl} = z_h(t) \hat{1}+ \sqrt{2\theta} \hat{d}^{\dagger}, \nonumber \\
&=& z_h(t) \sum_{n=0}^{\infty}|n><n|+ \sqrt{2\theta}\sum_{n=1}^{\infty}\sqrt{n-\nu}|n><n-1| .
 \label{sec2.18}
\eea
where $z_e(t)$ and $z_h(t)$ denote the center of mass coordinates of the quasi-electron and quasi-hole, respectively, in the complex notation.
The solution gives
\beq
[\hat{x}^1_e, \hat{x}^2_e]= i \theta (\hat{1}+ \nu |0><0|)
\label{sec2.20}
\eeq
at the location of the quasi-electron, and
\beq
[\hat{x}^1_h, \hat{x}^2_h]= -i \theta (\hat{1}- \nu |0><0|),
\label{sec2.21}
\eeq
at the location of the quasi-hole, so that the constraint(\ref{constraint of the 2 matrix mdel}) is satisfied where the exciton is located.  On the other locations there is no surplus or deficit of area, so that the coordinates are taken to be either $(\hat{z}_e)_{cl}$ or $(\hat{z}_h)_{cl}$ with $\nu=0$

In the above description we have repeatedly specified the locations where the
constraint is satisfied, saying "at the location of the quasi-electron, or at
the location of the quasi-hole, or on the other locations".  The reason for
this specification is as follows: In the contineous limit, the constraint
holds separately for each portion $(y^1, y^2)$ of the fluid.  That is, the
constraint holds at each location separately.  In the discretised version
of using matrices, however, the concept of the location becomes obscure. To
overcome this situation, we introduce into the solution, the center of mass
coordinates, $z_{e}(t)$ and $z_{h}(t) $, proportional to the unit matrix
$\hat{1}$.  These center of mass coordinates play the role of $y$ in the
contineous limit.  Therefore, we require the constraint condition to hold
separately for different values of these center of mass coordinates. We
consider the exciton solution in which the center of the mass coordinates
for the quasi-electron (location of the surplus of area) and the quasi-hole
(location of the deficit of area) are separated sufficiently, so that it is
also reasonable to require that at the location of the quasi-particle, the
constraint is satisfied by the electron part only, while at the location of
the quasi-hole, it is satisfied by the hole part only.

So far we have considered the Lagrangian $L_{0}$ which is the sum of Lorentz force term under the constant magnetic field and the constraint from the Pauli principle.  As is well known Lorentz force acts  perpendicular to the direction of motion, so that it does not contribute to the energy.  Therefore, if the constraint condition  is fulfilled, the Hamiltonian $H_{0}$ corresponding to the original $L_{0}$ vanishes;
\beq
H_{0} = 0.
\label{sec2.22}
\eeq
This vanishingness of the Hamiltonian is convenient to study electrons in the lowest Landau level, but to  study the excited states or the fractionally filling states, it is necessary to include the Landau level excitation energies as in Eqs.(\ref{sec1.11}) and (\ref{sec1.12}).
There are two ways to include such excitation modes.  The first standard way is to add the kinetic energy of the 2 dimensional electrons and holes by
\beq
L_{1} (2d) = \frac{1}{2} m_e Tr(\dot{\hat{x}}^a_e)^2 + \frac{1}{2} m_h Tr(\dot{\hat{x}}^a_h)^2 + (\mbox{ Coulomb potentials}),
\label{sec2.23}
\eeq
which we call the "2d" description.  The second way is to introduce the two dimensional potentials by
\beq
L_{1} (1d) = \frac{1}{4m_e}(eB)^2Tr(\hat{x}^a_e)^2 + \frac{1}{4m_h}(eB)^2Tr(\hat{x}^a_h)^2+(\mbox{ Coulomb potentials}),
\label{sec2.24}
\eeq
which we call the "1d" description.

Next, we discuss the meaning of and the relationship between these two different descriptions.
In the following we consider the quantum theory, in which we describes  the quantum operators in terms of capital letters.  Using this convention, the conjugate momenta $\hat{P}^a_e$ and $\hat{P}^a_h$ in the first description are given by
\bea
m_e \dot{\hat{X}}^a_e&=&\hat{P}^a_e -\frac{1}{2}eB\epsilon_{ab}\hat{X}^b_e, \nonumber \\
m_h \dot{\hat{X}}^a_h&=&\hat{P}^a_h +\frac{1}{2}eB\epsilon_{ab}\hat{X}^b_h,
\label{2.25}
\eea
and the Hamiltonian for this system becomes
\beq
H(2d)= \frac{1}{2m_e} Tr \left( \hat{P}^a_e -\frac{1}{2}eB\epsilon_{ab}\hat{X}^b_e  \right)^2 + \frac{1}{2m_h} \left( \hat{P}^a_h +\frac{1}{2}eB\epsilon_{ab}\hat{X}^b_h  \right)^2 + (\mbox{Coulomb potentials}).
\label{2.26}
\eeq
Here, we introduce the creation and annihilation operators by
\bea
\hat{A}_e &\equiv&\frac{1}{\sqrt{2}} \left[ \frac{i}{\sqrt{eB}} \left( \hat{P}_e^1-i\hat{P}_e^2 \right)+\frac{\sqrt{eB}}{2} \left( \hat{X}_e^1-i\hat{X}_e^2 \right) \right],
\nonumber \\
 \hat{A}^{\dagger}_e &\equiv&
          \frac{1}{\sqrt{2}} \left[ \frac{-i}{\sqrt{eB}} \left( \hat{P}_e^1+i\hat{P}_e^2 \right)
          +\frac{\sqrt{eB}}{2} \left( \hat{X}_e^1+i\hat{X}_e^2 \right) \right].
\label{new65}
\eea
and 
\bea
\hat{A}_h &\equiv&
           \frac{1}{\sqrt{2}} \left[ \frac{i}{\sqrt{eB}} \left( \hat{P}_h^1+i\hat{P}_h^2 \right)
          + \frac{\sqrt{eB}}{2} \left( \hat{X}_h^1+i\hat{X}_h^2 \right) \right]
\nonumber \\
 \hat{A}^{\dagger}_h &\equiv&
          \frac{1}{\sqrt{2}} \left[ \frac{-i}{\sqrt{eB}} \left( \hat{P}_h^1-i\hat{P}_h^2 \right)
          + \frac{\sqrt{eB}}{2} \left( \hat{X}_h^1-i\hat{X}_h^2 \right) \right]
\label{new66}
\eea
They satisfy the following quantum commutation relations,
\bea
\mathrm{[[}(\hat{A}_e)_{mn},(\hat{A}_e^{\dagger})_{n^{'}m^{'}}
\mathrm{]]}
    &=&\delta_{mm^{'}}\delta_{nn^{'}}
\nonumber\\
\mathrm{[[}(\hat{A}_h)_{mn},(\hat{A}_h^{\dagger})_{n^{'}m^{'}}
\mathrm{]]}
    &=&\delta_{mm^{'}}\delta_{nn^{'}},
 \label{new67}
\eea
based on  
\beq
 \mathrm{[[}(\hat{X}_{e(h)})_{nm},(\hat{P}_{e(h)})_{m^{'}n^{'}}
 \mathrm{]]}
 @=i\delta_{nn^{'}}\delta_{mm^{'}}. 
 \label{new68}
\eeq
Here the quantum commutator is specified by $[[O, O^{'}]]$, in order not to be confused with the commutator in the sense of the matrices, following ~\cite{Polychronakos}. The latter commutator is considered to be that in the "isospace", or in the "space of flavors".  
Then, the Hamiltonian becomes
\beq
 \hat{H}(2d)=\omega_eTr(\hat{A}_e^\dagger\hat{A}_e)
         +\omega_hTr(\hat{A}_h^\dagger\hat{A}_h) + (\mbox{zero point energy+ Coulomb potentials}),
 \label{new69}
\eeq
which is a reasonable one giving  Landau levels with proper cyclotron frequencies given in the previous section.

Now, we come to the second description.  In this description, $\hat{x}^2_e$ and  $\hat{x}^2_h$ are the coordinates, but can be considered as the conjugate momenta of  $\hat{x}^1_e$ and  $\hat{x}^1_h$, respectively.  Denoting $\hat{X}_e \equiv \hat{X}^1_e$ and  $\hat{X}_h \equiv \hat{X}^1_h $, the conjugate momenta are given by
\bea
\hat{P}_e &=& \frac{eB}{2}(\hat{X}_e^2), \nonumber \\
\hat{P}_h &=& -\frac{eB}{2}(\hat{X}_h^2). 
\label{2.32}
\eea
Then, the Hamiltonian in this description becomes the one dimensional harmonic oscillators:
\bea
H(1d)&=&Tr\left( \frac{1}{m_e} \hat{P_e}^2 + \frac{(eB)^2}{4m_e} \hat{X_e}^2 \right) + Tr\left( \frac{1}{m_h} \hat{P_h}^2 + \frac{(eB)^2}{4m_h} \hat{X_h}^2 \right) \nonumber \\
&+& (\mbox{Coulomb potentials}).
\label{2.33}
\eea
If we introduce the creation and annihilation operators by
\bea
\hat{A}_e &\equiv& \frac{1}{2} \left[ \sqrt{eB} \left(\hat{X}^1_e+i\hat{X}^2_e \right) \right] = \frac{1}{2} \left[ \sqrt{eB} \left(\hat{X}_e+i\frac{2}{eB} \hat{P}_e \right) \right], \nonumber \\
\hat{A}^{\dagger}_e &\equiv& \frac{1}{2} \left[ \sqrt{eB} \left( \hat{X}^1_e-i\hat{X}^2_e \right) \right] = \frac{1}{2} \left[ \sqrt{eB} \left( \hat{X}_e-i\frac{2}{eB}\hat{P}_e \right) \right],
\label{2.34}
\eea
and
\bea
\hat{A}_h &\equiv& \frac{1}{2} \left[ \sqrt{eB} \left( \hat{X}^1_h-i\hat{X}^2_h \right) \right] = \frac{1}{2} \left[ \sqrt{eB} \left( \hat{X}_h+i\frac{2}{eB}\hat{P}_h \right) \right], 
\nonumber \\ 
\hat{A}^{\dagger}_h  &\equiv& \frac{1}{2} \left[ \sqrt{eB} \left( \hat{X}^1_h+i\hat{X}^2_h \right) \right] = \frac{1}{2} \left[ \sqrt{eB} \left( \hat{X}_h-i\frac{2}{eB}\hat{P}_h \right) \right].
\label{2.35}
\eea
Then, the Hamiltonian in this second description reads
\bea
H(1d) &=& \omega_e Tr \left( \hat{A}^{\dagger}_e \hat{A}_e \right) + \omega_h Tr \left( \hat{A}^{\dagger}_h \hat{A}_h \right)  \nonumber \\
&+& (\mbox{zero point energy+ Coulomb potentials}).
\label{2.36}
\eea
Therefore, the final Hamiltonians and their eigen-values of these two descriptions are shown to be equivalent at the quantum mechanics.

We have said "equivalent" here, but as one can understand well, the 1d system
and the 2d system are different dynamical systems, having different degrees
of freedom.  Only thing we have claimed here is that these 1d and 2d
systems play effectively the equivalent roles in the calculation of the
energy levels quantum mechanically.

To understand the physical correspondence between these two descriptions, it is better to examine both pictures semi-classically.  Let us simplify the infinite matrix to a single component one including only the electron, and study the semi-classical motions without Coulomb potentials. 
Here, the simplified Hamiltonian $H_s$ of the "2d" picture and "1d" picture are respectively,
\bea
H_s(2d)&=&\frac{1}{2} m_e \sum_{a=1,2} (\dot{x}^a)^2 = \frac{1}{2m_e} \sum_{a=1,2} \left( p^a-\frac{eB}{2}\epsilon_{ab}(x^b-x^b_{cm}) \right)^2,
\label{2.37}\\ 
H_s(1d)&=&\frac{(eB)^2}{4m_e} \sum_{a=1,2} (x^a-x^a_{cm})^2 = \frac{(eB)^2}{4m_e} (x-x_{cm})^2 + \frac{1}{m_e} (p-p_{cm})^2,
\label{2.38}
\eea
where the variables with the suffix "cm" denote the values of them at the center of the cyclotron motion. 
Let us start with the "2d" picture. It gives a cyclotron motion under the constant magnetic field, namely,
\bea
x^1&=&x^1_{cm} + r_{2d} \sin (\omega_e t), \nonumber \\
x^2&=&x^2_{cm} + r_{2d} \cos (\omega_e t).
\label{2.39}
\eea
Its energy is
\beq
H_s(2d)=\frac{1}{2} m_e (r_{2d})^2\omega_e^2 = \omega_e (n+1/2),
\label{2.40}
\eeq
where the quantization can be considered as the quantization of the angular momentum,
\beq
M=m_e (r_{2d})^2 \omega_e= 2n+1 ~~(n: \mbox{integer}).
\label{2.41}
\eeq
In the second "1d" picture, it gives a one dimensional harmonic oscillator of a point particle with mass $m_e/2$ and spring constant $(eB)^2/2m_e$, namely
\beq
x-x_{cm}= r_{1d} \sin (\omega_e t),
\label{2.42}
\eeq
having the same frequency as in the "2d" picture.
The corresponding momentum is given by
\beq
x^2-x^2_{cm} \equiv \frac{2}{eB}(p-p_{cm})=r_{1d}\cos (\omega_e t),
\label{2.43}
\eeq
so that we have the circular motion in the  $(x^1, x^2)$ space, or equivalently the elliptic motion in the one dimensional "phase space" of $(x, p)$.
The energy of the "1d" picture reads
\beq
H_s(1d)=\frac{1}{4}m_e (r_{1d})^2 \omega_e^2 =\omega_e (n+1/2),
\label{2.44}
\eeq
where the latter identity is considered to be the quantization for the phase space area, namely,
\beq
\pi (r_{1d})^2=2\pi (r_{2d})^2= 2\pi (2n+1) (eB)^{-1}.
\label{2.45}
\eeq
Since the expressions of energies are identical in equation (\ref{2.40})and (\ref{2.44}), two radii $r_{1d}$ and $r_{2d}$ are related as 
\beq
r_{1d}=\sqrt{2} r_{2d}.
\eeq

In conclusion, there are two equivalent ("1d" and "2d" ) pictures to describe the cyclotron motion under the constant magnetic field, and the relationship between these two pictures are as follows: The phase space trajectory of the "1d" picture is equivalent to the real space trajectory of the "2d" picture, and the quantization of the phase space area in the "1d" picture corresponds to the quantization of the angular momentum in the "2d" picture.
\section{Physical States of the Quantized Matrix Model}
In this section we will study what kinds of physical states are allowed, after quantizing the two component matrix model in the second description, {\it i.e.}, in the "1d" picture, in the study of the quantum Hall effect.

In case of having a classical solution such as the exciton solution, we expand the fields around the classical solutions $(\hat{x}_e^a)_{cl}$ and  $(\hat{x}_h^a)_{cl}$, and the quantum fluctuations denoted by the capital letters $\hat{X}_e^a$ and $\hat{X}_h^a$ as: 
\bea
\hat{x}_e^a&=&(\hat{x}_e^a)_{cl}+\hat{X}_e^a\nonumber\\
\hat{x}_h^a&=&(\hat{x}_h^a)_{cl}+\hat{X}_h^a.
\label{new1}
\eea
The canonical commutation relations of the "1d" picture can be written as
\beq
[[(\hat{X_e^1})_{mn},(\hat{X_e^2})_{n^{'}m^{'}}\mathrm{]]}
=i\hbar\frac{2}{eB} \delta_{mm^{'}} \delta_{nn^{'}},  
\label{3.48}
\eeq
\beq
[[(\hat{X_h^1})_{mn},(\hat{X_h^2})_{n^{'}m^{'}}\mathrm{]]}
=i\hbar\frac{2}{eB} \delta_{mm{'}} \delta_{nn^{'}},
\label{new2}
\eeq
and the conjugate momenta $\hat{P}_e$ and $\hat{P}_h$ of $\hat{X}_e (\equiv \hat{X}^1_e)$ and $\hat{X}_h (\equiv \hat{X}^1_h)$ are
\bea
(P_e)_{nm} &\equiv& \frac{eB}{2}(\hat{X}_e^2)_{nm}
     =-i\frac{\delta}{\delta (\hat{X}_e)_{mn}}, 
\label{3.50}
\\
(P_h)_{nm} &\equiv& -\frac{eB}{2}(\hat{X}_h^2)_{nm}
     =-i\frac{\delta}{\delta (\hat{X}_h)_{mn}}. 
\label{new3}
\eea

In this case the constraint condition of the two component matrix model given in Eq.(\ref{constraint of the 2 matrix mdel}), can be naturally decomposed into the purely classical part, 
\beq
[(\hat{x}_e^1)_{cl},(\hat{x}_e^2)_{cl}]-[(\hat{x}_h^1)_{cl},(\hat{x}_h^2)_{cl}]=i\theta\hat{\bf{1}},
\label{3.52}
\eeq
the mixed part of classical and quantum effects,
\beq
\mathrm{[}(\hat{x}_e^1)_{cl}, \hat{X}_e^2\mathrm{]}
+\mathrm{[}\hat{X}_e^1,(\hat{x}_e^2)_{cl}\mathrm{]}
-\mathrm{[}(\hat{x}_h^1)_{cl}, \hat{X}_h^2\mathrm{]}
-\mathrm{[}\hat{X}_h^1,(\hat{x}_h^2)_{cl}\mathrm{]} =0,
 \label{3.53}
\eeq
and the purely quantum part,
\beq
\mathrm{[}\hat{X}_e^1, \hat{X}_e^2\mathrm{]}
-\mathrm{[}\hat{X}_h^1, \hat{X}_h^2\mathrm{]}=0.
\label{3.54}
\eeq



Here, it is better to comment on the symmetry of our system for both
the hydrodynamical model and the matrix model of the quantum Hall effect.
  In the contineous limit of $\theta \rightarrow 0$, 
our system becomes the fluid
dynamics, having the gauge symmetry called the "area-preserving
diffeomorphism": $(x^1_{e}, x^2_{e}) \rightarrow ({x^1}^\prime _{e},
 {x^2}^\prime _{e})$ and
$(x^1_{h}, x^2_{h}) \rightarrow ({x^1}^\prime _{h}, {x^2}^\prime _{h})$, having the following
invariance for the "Poisson brackets",
$$ \frac{\partial(x^1_{e}, x^2_{e})}{\partial(y^1,
y^2)}-\frac{\partial(x^1_{h}, x^2_{h})}{\partial(y^1,
y^2)}=\frac{\partial({x^1}^\prime _{e}, {x^2}^\prime _{e})}{\partial(y^1,
y^2)}-\frac{\partial({x^1}^\prime _{h}, {x^2}^\prime _{h})}
{\partial(y^1, y^2)}$$.  Under
this gauge transformation, the constraint condition is invariant.
When moving to the matrix model of the quantum Hall effect, the gauge
symmetry of area preserving differomorphism becomes the unitary
transformation for the matrices.  This is easily understood if we recognize
that $\theta$ plays the role of $\hbar$ in quantum mechanics and that the
limit $\theta \rightarrow 0$ corresponds to the classical limit of quantum
mechanics.  Area-preserving diffeomorphism in the quantum Hall fluid
corresponds to the canonical transformation preserving the "phase space
area" in quantum mechanics.  The canonical transformation becomes the
unitary transformation in quantum mechanics, so that the matrix version of
our symmetry is the invariance under the $SU(\infty)$ transformation,
namely,
\begin{eqnarray}
\hat{x_{e}} &\to& \hat{x_{e}}'=U\hat{x_{e}} U^{-1}, \nonumber
\\
\hat{x_{h}} &\to& \hat{x_{h}}'=U\hat{x_{h}} U^{-1}, \nonumber
\\
\hat{A} &\to& \hat{A}'=U \hat{A}U^{-1},
\end{eqnarray}
where $U$ is an arbitrary $SU(\infty)$ matrix. The invariance of our two
matrix model under this $SU(\infty)$ symmety can be checked directly from
the original action in Eq. (\ref{e12}).  In this paper this symmetry is somtimes
called "gauge symetry" or "area-preserving diffeomorphisms".
The "gauge fixing" stated above is the fixing of the arbitrariness with
respect to the $SU(\infty)$ symmetry.


The vanishingness of the purely classical part is guaranteed by the classical constraint condition.  Therefore, the remaining sum of the  mixed part of classical and quantum effects and the purely quantum part vanishes.  The vanishingness of the mixed part can be chosen as gauge condition (the background gauge condition).  To quantize with this gauge condition of the fluctuations around the classical solution (such as our exciton solution), we have to extract properly the collective coordinates ({\it i.e.}the center of mass coordinates and others) of the classical solution from the quantum fluctuations \cite{Sakita's book}.  We leave this problem for a future study, and naively consider the quantization procedure here in order to obtain the allowed physical states.

Then, the quantum constraint operator is equal to the constraint operator of the purely quantum part given by,
\bea
\hat{G}_{mn}&=&([\hat{X}_e^1,\hat{X}_e^2])_{mn}
             -([\hat{X}_h^1,\hat{X}_h^2])_{mn} \nonumber\\
&=& \frac{2}{eB} \{
(\hat{X}_e)_{mk}(\hat{P}_e)_{nk}-(\hat{P}_e)_{km}(\hat{X}_e)_{kn}
     \nonumber\\
  & &-(\hat{X}_h)_{mk}(\hat{P}_h)_{nk}+(\hat{P}_h)_{km}(\hat{X}_h)_{kn}
     \},
\label{new10}
\eea
the discretized version of the generators of the "area-preserving diffeomorphisms" in the quantum Hall fluid. The "area" means the area occupied by the electron fluid minus that of the hole fluid.  Now, the physical states $|phys>>$ in our quantized two component matrix model should satisfy the following constraint:
\beq
\hat{G}_{mn}|phys>>=0.
  \label{new30}
\eeq
The physical states of the two component matrix model are spanned without the Coulomb interactions by the creation operators of electron and holes in the "1d" picture, namely, $\hat{A}^{\dagger}_e$ and $\hat{A}^{\dagger}_h$ in Eqs. (\ref{2.34}) and (\ref{2.35}) defined in the last section.  In terms of these creation operators the constraint operator $\hat{G}_{mn}$ can be written as
\bea
 \hat{G}_{mn}=i\left(\frac{4}{eB}\right) \displaystyle{\sum_k}
    \{(\hat{A}_e^\dagger)_{kn}(\hat{A}_e)_{mk}
   &-&(\hat{A}_e^\dagger)_{mk}(\hat{A}_e)_{kn}
 \nonumber\\
+ (\hat{A}_h^\dagger )_{kn}(\hat{A}_h)_{mk}
   &-&(\hat{A}_h^\dagger )_{mk}(\hat{A}_h)_{kn}\}, \\
=i\left(\frac{4}{eB}\right)\displaystyle{\sum_k}
   \{
   (\hat{A}_e^\dagger)_{kn}\frac{\delta}{\delta (\hat{A}_e^\dagger)_{km}}
&-&(\hat{A}_e^\dagger)_{mk}\frac{\delta}{\delta (\hat{A}_e^\dagger)_{nk}}
 \nonumber\\
   +(\hat{A}_h^\dagger)_{kn}\frac{\delta}{\delta (\hat{A}_h^\dagger)_{km}}
&-&(\hat{A}_h^\dagger)_{mk}\frac{\delta}{\delta (\hat{A}_h^\dagger)_{nk}}
   \},
\label{new42}
\eea
where the normal ordering is taken for the constraint operator.

The meaning of the constraint operator $\hat{G}_{mn}$ is the following: It creates the difference of two states obtained by performing the following two replacements on the creation operators defining a state. The first one is obtained  by replacing the (k, m) component of the creation operators by their (k, n) component for every k-th row with a common coefficient.  The second one replaces the (n, k) component of the  creation operators by their (m, k) component for every k-th column with a common coefficient. 
Then, we can prove the following formula for an integer positive power of $p$:
\bea
\hat{G}_{mn} (\hat{A}^{\dagger p})_{ij}=\delta_{jm}(\hat{A}^{\dagger^p})_{in} - \delta_{in}(\hat{A}^{\dagger p})_{mj}.
\label{3.59}
\eea
Therefore, the above meaning of how the constraint operator works is valid also for the cluster of $\hat{A}^p$.  When the constraint    operator acts on a certain determinant, it replaces m-th column by its n-th column, or n-th row by its m-th row, leading the determinant to vanish.  Also the constraint operator acting on the trace $Tr(\hat{A}^{\dagger p})$ leads to vanish.

Therefore, the following state $|\Psi>>\equiv|(m; p_1, p_2, \dots; q); (n; p^{'}_1, p^{'}_2, \dots; q^{'})>>$ can be proved to be a physical state:
\bea
 & &|\Psi>>\equiv
 |(m; p_1, p_2, \dots; q); (n; p^{'}_1, p^{'}_2, \dots; q^{'})>> 
\nonumber \\
&\equiv& Tr\left(\hat{A}^{\dagger}_h\right)^q
         \displaystyle{\sum_{\{i\}}}
         \left(\epsilon_{i_1i_2...i_N }(\hat{A}^{\dagger}_h)^{p_1}_{1,i_1}
         (\hat{A}^{\dagger}_h)^{p_2}_{2,i_2}...\right)^{m}\nonumber\\
&\times&  Tr\left(\hat{A}^{\dagger}_e\right)^{q^{'}}
         \displaystyle{\sum_{\{i\}}}
         \left(\epsilon_{i_1i_2...i_N}(\hat{A}^\dagger_e)^{p^{'}_1}_{1,i_1}
         (\hat{A}^\dagger_e)^{p^{'}_2}_{2,i_2}...\right)^{n}|0>>,
  \label{new48}
\eea
where all the indices $(m; p_1, p_2, \dots; q); (n; p^{'}_1, p^{'}_2, \dots; q^{'})$ are positive integers, and $|0>>$ denotes the ground state.
If the Coulomb potentials can be ignored and if $\omega_e \gg \omega_h$ holds or we do not care about the hole energy (the same assumption which is imposed in the introduction to get the Laughlin states), the following state is the special example of the above mentioned physical states:
\bea
& &|\Psi_{LL}>>_m\equiv|(m; 1, 2,  ...; 0); (0; 0, 0, ...; 0)>>
\nonumber \\
&\equiv&
\displaystyle{\sum_{\{i,j\}}}
\left(\epsilon_{i_1i_2i_3...}\epsilon_{j_1j_2j_3...}
     (\hat{A}^{\dagger}_h)^{0}_{i_1,j_1}
  (\hat{A}^{\dagger}_h)^{1}_{i_2,j_2} (\hat{A}^{\dagger}_h)^{2}_{i_3,j_3}...
\right)^{m}|0>>.
\label{3.61}
\eea
Although the nature of this state is not clear now, we are tempting to expect that this state plays a similar important role as is done by the Laughlin state, in the matrix formulation of the quantum Hall effect.  Therefore we call this the Laughlin like states (LL) denoted by $|\Psi_{LL}>>_m$.

To understand the physical meaning of the states, the coordinate representation of the physical state or the wave function of it is useful.  This direction is already studied using the coherent state methods in the finite matrix model \cite{Sakita}.
In our case, we can express the two sets of the creation and annihilation operators, $(\hat{A}_e, \hat{A}^{\dagger}_e)$ and $(\hat{A}_h, \hat{A}^{\dagger}_h)$ by the real coordinates $\hat{X}\equiv\hat{X}^1_e$ and $\hat{Y}\equiv\hat{X}^2_h$, or by the complex coordinates $\hat{Z}\equiv \hat{X}+i\hat{Y}$ and $\hat{\bar{Z}}\equiv \hat{X}-i\hat{Y}$ as follows:
\bea
(\hat{A}_e)_{mn}&=&\frac{1}{2}\sqrt{eB}
  \left((\hat{X})_{mn}+\frac{2}{eB}\frac{\delta}{\delta(\hat{X})_{nm}}\right)
\nonumber\\
(\hat{A}_e^\dagger)_{nm}&=&\frac{1}{2}\sqrt{eB}
 \left((\hat{X})_{nm}
 -\frac{2}{eB}\frac{\delta}{\delta(\hat{X})_{mn}}\right)
\label{new84}
\\
(\hat{A}_h)_{mn}&=&\frac{1}{2}\sqrt{eB}
  \left(\frac{2}{eB}(-i)\frac{\delta}{\delta(\hat{Y})_{nm}}-i(\hat{Y})_{mn}
  \right)
\nonumber\\
(\hat{A}_h^\dagger)_{nm}&=&\frac{1}{2}\sqrt{eB}
 \left(\frac{2}{eB}(-i)\frac{\delta}{\delta(\hat{Y})_{mn}}+i(\hat{Y})_{nm}
 \right)
\label{new86}
\eea
\bea
\frac{1}{\sqrt{2}}(\hat{A}_e-\hat{A}_h)_{mn}
   &\equiv&\frac{1}{\sqrt{2}}\left(
    \frac{1}{\sqrt{eB}}2\frac{\delta}{\delta(\hat{\bar{Z}})_{nm}}
   +\frac{1}{2}\sqrt{eB}(\hat{Z})_{mn}\right)
    \equiv\hat{A}_{-}
\nonumber\\
\frac{1}{\sqrt{2}}(\hat{A}_e^\dagger-\hat{A}_h^\dagger)_{nm}
   &\equiv&\frac{1}{\sqrt{2}}\left(
    \frac{1}{\sqrt{eB}}(-1)\times 2\frac{\delta}{\delta(\hat{Z})_{mn}}
   +\frac{1}{2}\sqrt{eB}(\hat{\bar{Z}})_{nm}\right)
    \equiv\hat{A}_{-}^\dagger
\nonumber\\
\label{new88}\\
\frac{1}{\sqrt{2}}(\hat{A}_e+\hat{A}_h)_{mn}
   &=&\frac{1}{\sqrt{2}}\left(
    \frac{1}{\sqrt{eB}}2\frac{\delta}{\delta(\hat{Z})_{nm}}
   +\frac{1}{2}\sqrt{eB}(\hat{\bar{Z}})_{mn}\right)
    \equiv\hat{A}_{+}
\nonumber\\
\frac{1}{\sqrt{2}}(\hat{A}_e^\dagger+\hat{A}_h^\dagger)_{nm}
   &=&\frac{1}{\sqrt{2}}\left(
    \frac{1}{\sqrt{eB}}(-1)\times 2\frac{\delta}{\delta(\hat{\bar{Z}})_{mn}}
   +\frac{1}{2}\sqrt{eB}(\hat{Z})_{nm}\right)
    \equiv\hat{A}_{+}^\dagger
\nonumber\\
\label{new89}
\eea
Then, the wave functions of the physical states mentioned above can be obtained by acting successively the creation operators of electrons and holes in the real or the complex representations, on the ground state $|0>>$ given by
\beq
|0>> \propto \exp \{-\frac{1}{4}(eB) Tr(\hat{Z}\hat{\bar{Z}}) \}.
\label{3.64}
\eeq
The structure of the physical states is identical to the Laughlin's case which we can find in\cite{Laughlin1}, but having some  differences, namely, the complex coordinates are replaced by the matrices, as well as the antisymmetrization of the wave function to grantee the Fermi statistics is naturally replaced by the physical state condition or by the invariance under the discrete version of the area preserving diffeomorphisms.
\section{Exciton Energy}
In this section, we come back to the original problem of the exciton, the neutral excitation of quantm Hall effect, having  both quasi-electron with charge  $-\nu e$  and quasi-hole with charge  $\nu e$.  Based on the discussion given in the last section on the physical states, we will study the energy of the exciton under the given background quantum state.
As such a background state we examine a rather general physical state of $|\Psi>>=|(m; p_1, p_2, \dots; q); (n; p^{'}_1, p^{'}_2, \dots; q^{'})>>$ given in Eq.(\ref{new48}) and its special case of $|\Psi_{LL}>>_m$, the Laughlin-like state.

Energy of the exciton consists of self-energies of both quasi-electron and quasi-hole, and of the attractive Coulomb potential energy between them. We do not include the Coulomb self-energies of the individual quasi-electron and quasi-hole as usual, since they are probably negligible small compared to their kinetic energies in the strong magnetic field $B$.
Let us start with the following Hamiltonian adequate for the evaluation of the exciton energy, namely,
\bea
H_{exc}&=&\frac{(e^{(0)}B)^2}{4m^{(0)}_{e}} Tr \sum_{a=1,2} \left( \hat{x_e}^a-(\hat{x_e}^a)_{cm} \right)^2 + \frac{(e^{(0)}B)^2}{4m^{(0)}_{h}} Tr \sum_{a=1,2} \left( \hat{x_h}^a-(\hat{x_h}^a)_{cm} \right)^2
\nonumber \\
& & - \frac{(e^{(0)})^2}{4\pi\varepsilon^{(0)}} Tr \left( \sum_{a=1,2}\left( \hat{x}_e^a-\hat{x}_h^a \right)^2 \right)^{-1/2}.
\label{4.65}
\eea
We consider the initial parameters of charge $e^{(0)}$, masses of electron and hole,  $m^{(0)}_e$ and  $m^{(0)}_h$, and dielectric constant $\varepsilon^{(0)}$ to be something like the "bare" quantities. They will be adjusted later by the "renormalization" in order to reproduce the physics of the exciton.

As was discussed before in equation (\ref{new1}), the coordinates are described as the sum of the classical configurations and the quantum fluctuations, or in terms of the complex variables,
\bea
\hat{z}_e &=&(\hat{z}_e)_{cl} + \frac{2}{\sqrt{eB}}\hat{A}_e,
\nonumber \\
\hat{z}_h &=&(\hat{z}_h)_{cl} + \frac{2}{\sqrt{eB}}\hat{A}^{\dagger}_h,
\label{4.67}
\eea
where $(\hat{z}_e)_{cl}$ and $(\hat{z}_h)_{cl}$ are those given in Eqs.(\ref{sec2.17}) and (\ref{sec2.18}), derived from the constraint condition giving the deficit and the surplus of "area" $q=2\pi\theta^{(0)}\nu$.  Here, we start with the "bare" variable $\theta^{(0)}$, but consider $\nu$ as a physical parameter from the beginning, not to be renormalized. 
Then, in the lowest perturbation theory in quantum mechanics, the exciton energy under the given background quantum state $|\Psi>>$ can be expressed by the expectation value:
\beq
E_{exc} = \frac{<<\Psi| H_{exc} |\Psi>>}{<<\Psi|\Psi>>}.
\label{4.68}
\eeq 

The corrections of self-energies will appear in the cyclotron radius squared.  There are two kinds of corrections: One kind of corrections originates from the noncommutativity of space (Pauli principle) at the individual location of quasi-electron or quasi-hole, and is expanded in $\theta^{(0)}$, while the other one comes from the quantum corrections from the background states and is expanded in $1/e^{(0)}B$, so that we have
\bea
Tr \sum_{a=1,2} 
\left( \hat{x_e}^a-(\hat{x}_e^a)_{cm}\right)^2 
   &=&| z_e-(z_e)_{cm}|^2 (Tr\hat{1})+ \theta^{(0)}Tr(\hat{C}_{NC(e)})
      + \frac{2}{e^{(0)}B}Tr(\hat{C}_{Q(e)}),
\nonumber\\
\label{4.69}
\\
Tr \sum_{a=1,2} \left( \hat{x}_h^a-(\hat{x}_h^a)_{cm} \right)^2&=&| z_h-(z_h)_{cm}|^2 (Tr\hat{1})+ \theta^{(0)} Tr(\hat{C}_{NC(h)}) + \frac{2}{e^{(0)}B}Tr(\hat{C}_{Q(h)}).
\nonumber\\
\label{4.70}
\eea
Here the noncommutative corrections $\hat{C}_{NC(e)}$ and $\hat{C}_{NC(h)}$, and the quantum corrections $\hat{C}_{Q(e)}$ and $\hat{C}_{Q(h)}$ are given by
\bea
Tr(\hat{C}_{NC(e)}) &\equiv& \lim_{N \rightarrow \infty}\sum_{n=0}^{N}<n| \hat{b}\hat{b}^{\dagger}+ \hat{b}^{\dagger}\hat{b} |n>,
\label{4.71}
\\
Tr(\hat{C}_{NC(h)}) &\equiv&
\lim_{N \rightarrow \infty}\sum_{n=0}^{N}<n| \hat{d}\hat{d}^{\dagger}+ \hat{d}^{\dagger}\hat{d} |n>,
\label{4.72}
\eea
and
\bea
Tr(\hat{C}_{Q(e)}) &\equiv&  Tr\left(\hat{A_e}\hat{A_e}^{\dagger}+\hat{A_e}^{\dagger}\hat{A_e}\right).
\label{4.73}
\\
Tr(\hat{C}_{Q(h)}) &\equiv& Tr\left(\hat{A_h}\hat{A_h}^{\dagger}+\hat{A_h}^{\dagger}\hat{A_h}\right). 
\label{4.74}
\eea

To estimate the Coulomb potential energy, we further expand the matrix version of the distance squared between quasi-electron and quasi-hole around the mean distance squared between them, namely, $|z_e-z_h|^{2}$, and keep the lowest order corrections.  Then, we have
\bea
& &Tr \left( \sum_{a=1,2}
\left( \hat{x}_e^a-\hat{x}_h^a \right)^2 \right)^{-1/2}
\nonumber \\
&=& \frac{1}{|z_e-z_h|}  \left\{
Tr (\hat{1}) + \frac{(-1/2)}{|z_e-z_h|^2}\left( \theta^{(0)}Tr\left(\hat{C}_{NC(e\&h)}\right) + \frac{2}{e^{(0)}B}Tr(\hat{C}_{Q(e)}+\hat{C}_{Q(h)}) \right) \right\},
\nonumber\\
\label{4.75}
\eea
where the noncommutative correction $Tr
\left(\hat{C}_{NC(e\&h)}\right)$ given by
\bea
Tr\left(\hat{C}_{NC(e\&h)}
\right)
=\lim_{N \rightarrow \infty} \sum_{n=0}^{N}<n|
  \{
  (\hat{b}\hat{b}^{\dagger}+\hat{b}^{\dagger}\hat{b})
 &+&(\hat{d}\hat{d}^{\dagger}+\hat{d}^{\dagger}\hat{d})
\nonumber \\
 -(\hat{b}\hat{d}+\hat{d}\hat{b})&-&(\hat{b}^{\dagger}\hat{d}^{\dagger}
 +\hat{d}^{\dagger}\hat{b}^{\dagger})
  \}|n>,
\label{4.76}
\eea
represents the noncommutativity of  space (Pauli principle) having the correlation between quasi-electron and quasi-hole locations.
While the quantum corrections in this case come from both quantum states of electrons and holes, namely, $Tr(\hat{C}_{Q(e)}+\hat{C}_{Q(h)})$.

In estimating the trace of the infinite-sized matrix we are using so far,  it may be possible to use a fancy technique of the $\zeta$ function regularization, using such as Riemann's and Epstein's $\zeta$ functions~\cite{Odintsov}.  Here we adopt a most naive method, however,  by introducing a cutoff $N$ on the size of the matrix, while keeping $N$ large but not infinite.  We denote the corresponding regularized trace as $Tr_N$, which is obtained by replacing the infinite sum $\sum_{n=0}^{\infty}$ in $Tr$ by the finite sum $\sum_{n=0}^{N}$.
Then, we have the following results:
For the noncommutativity corrections, we obtain
\bea
Tr_{N} (\hat{C}_{NC(e)}) &=&\left(\sum_{n=1}^{N}+\sum_{n=1}^{N+1}\right) (n+\nu) = (N+1)^2 + \nu (2N+1),
\label{77}
\\
Tr_{N} (\hat{C}_{NC(e)}) &=&\left(\sum_{n=1}^{N}+\sum_{n=1}^{N+1}\right) (n-\nu) = (N+1)^2 - \nu (2N+1),
\label{78}\\
Tr_{N} (\hat{C}_{NC(e\&h)}) &=& \left(\sum_{n=1}^{N}+\sum_{n=1}^{N+1}\right) (n+\nu)+ \left(\sum_{n=1}^{N}+\sum_{n=1}^{N+1}\right) (n-\nu) 
\nonumber \\
&-& \left(\sum_{n=1}^{N-1}+\sum_{n=1}^{N+1}\right) \left(\sqrt{(n+\frac{1}{2})^2 - (\nu+\frac{1}{2})^2}+\sqrt{(n+\frac{1}{2})^2 - (\nu-\frac{1}{2})^2}\right)
\nonumber \\
&=&(\nu^2+\frac{1}{4}) \left(\sum_{n=1}^{N-1} +\sum_{n=1}^{N+1} \right) \frac{1}{n+\frac{1}{2}} + \cdots 
\label{79}
\\
&\approx& 2 (\nu^2+\frac{1}{4}) \ln N,
\label{80}
\eea
where we have to note that in the third trace, $(N+1)^2$ terms cancel completely, and $\ln N$ contribution remains in the large $N$ limit. 
As for the quantum corrections, they consist of the number operators, so that in the lowest order we can easily obtain by using 
\bea
& & C^{(0)}_Q \equiv \frac{<<\Psi|Tr_N(\hat{C}_Q)||\Psi>>}{<<\Psi|\Psi>>}
\label{81}\\
&=&\sum_{modes} \left( 2\times(\mbox{number of excited quanta}) + 1 \right).
\label{82}
\eea
As an example, for a general state of $|\Psi>>=|(m; p_1, p_2, \dots; q); (n; p^{'}_1, p^{'}_2, \dots; q^{'})>>$we have
\bea
C^{(0)}_{Q(h)} &=& 2\left(m \sum_{i=1}^{N+1} p_i + (N+1) q \right) + (N+1)^2,
\label{83}\\
C^{(0)}_{Q(e)}&=& 2\left(n \sum_{i=1}^{N+1} p^{'}_i + (N+1) q^{'} \right) +(N+1)^2.
\label{84}
\eea
For the Laughlin-like state of $|\Psi_{LL}>>_m=|(m; 0, 1, 2, \dots; q); (0; 0, 0, \dots; 0)>>$, we have 
\bea
C^{(0)}_{Q(h)}(LL)&=&2m\sum_{n=1}^{N}n+(N+1)^2=mN(N+1)+(N+1)^2,
\label{85}
\\
C^{(0)}_{Q(e)}(LL)&=&(N+1)^2.
\label{86}
\eea

Now, we can write down the exciton energy in the large $N$ limit in the following:
\bea
& & H_{exc}\hfill
\nonumber \\
&=&\frac{(e^{(0)}B)^2}{4m^{(0)}_{e}} 
\left\{ (N+1) |z_e-(z_e)_{cm}|^2 + \theta^{(0)} 
    \left( (N+1)^2 +\nu(2N+1) \right) 
    +\frac{2}{e^{(0)}B}C^{(0)}_{Q(e)}
 \right\}
\nonumber \\
&+& \frac{(e^{(0)}B)^2}{4m^{(0)}_{h}}
   \left\{ (N+1) |z_h-(z_h)_{cm}|^2 + \theta^{(0)} 
   \left( (N+1)^2 - \nu(2N+1) \right) +\frac{2}{e^{(0)}B}C^{(0)}_{Q(h)} 
   \right\}  
\nonumber \\
&-&\frac{(e^{(0)})^2}{4\pi\varepsilon^{(0)}} \frac{1}{|z_e-z_h|}
\left\{ (N+1) - \frac{1}{|z_e-z_h|^2}
 \left(
 \theta^{(0)}(\nu^2+\frac{1}{4}) \ln N 
  +\left(\frac{C^{(0)}_{Q(e)}+C^{(0)}_{Q(h)}}{e^{(0)}B}\right)
   \right) 
  \right\}.
\nonumber\\
\label{87}
\eea
There are various infinities in this expression in the limit of $N \rightarrow \infty$, since we have not started with the correct normalizations for the infinite coordinate matrices. Therefore, we "renormalize", or in other words, simply change of the normalization for various "bare" parameters to reproduce the physics of the exciton.  Namely, we impose the following:
\bea
\frac{(e^{(0)})^2}{4m^{(0)}_e}(N+1) &\equiv& \frac{(\nu e)^2}{4m_{qe}},
\label{88}\\
\frac{(e^{(0)})^2}{4m^{(0)}_h}(N+1) &\equiv& \frac{(\nu e)^2}{4m_{qh}},
\label{89}\\
\theta^{(0)} (N+1) &\equiv& \theta,
\label{90}\\
\frac{(e^{(0)})^2}{4\pi\varepsilon^{(0)}} (N+1) &\equiv& \frac{(\nu e)^2}{4\pi\varepsilon},
\label{91}\\
\frac{e^{(0)}}{(N+1)}\equiv e,
\label{92}
\eea
where $e, m_{qe}, m_{qh}, \theta, \varepsilon$ are physical parameters, representing the unit charge, the effective mass of the quasi-electron, the effective mass of the quasi-hole, the noncommutativity parameter, and the effective dielectric constant of quasi-electron and quasi-hole in the quantum Hall medium, respectively.

In terms of these physical parameters, exciton energy can be written as
\bea
E_{exc}&=& 
 \frac{(\nu eB)^2}{4m_{qe}}
  \left(|z_e-(z_e)_{cm}|^2 + \theta  + \frac{2}{eB} C_{Q(e)}\right)
\nonumber \\
&+&
 \frac{(\nu eB)^2}{4m_{qh}}
  \left(|z_h-(z_h)_{cm}|^2+\theta+\frac{2}{eB}C_{Q(h)}\right)
\nonumber \\
&-&
 \frac{(\nu e)^2}{4\pi \varepsilon |z_e-z_h|}
  \left(1-\frac{1}{eB|z_e-z_h|^2}
   \left(C_{Q(e)} + C_{Q(h)}\right)+ \cdots
  \right).
\label{93}
\eea
The Coulomb potential term in the above equation can also be approximately given by the following expression:
\beq
-\frac{(\nu e)^2}{4\pi\varepsilon
 \left(|z_e-z_h|^2 + \frac{2}{eB}\left(C_{Q(e)}+C_{Q(h)}\right) \right)^{1/2}}.
\label{94}
\eeq
In the above expressions, $C_{Q(e~or~h)}$ is the renormalized coefficient representing the quantum corrections defined by
\beq
C_{Q(e~or~h)} \equiv \lim_{N \rightarrow \infty} \frac{C^{(0)}_{Q(e~or~h)}} {(N+1)^2}.
\label{95}
\eeq
For a general state $|\Psi>>$, we have
\bea
C_{Q(h)}&=&2 \lim_{N \rightarrow \infty}\frac{m\sum_{i}p_{i}}{(N+1)^2} + 1,
\label{96}
\\
C_{Q(e)}&=&2 \lim_{N \rightarrow \infty}\frac{n\sum_{i}p^{'}_{i}}{(N+1)^2} + 1.
\label{97}
\eea
For most of the states, $C_{Q(e~or~h)}=1$, meaning that only the zero point energies of the electron or the hole contribute to this number, but for the Laughlin like state, however, the number becomes $C_{Q(h)}(LL)=m+1$ and $C_{Q(e)}(LL)=1$.

What happens physically for the exciton energy?  
As was pointed out previously, all the corrections coming from both the noncommutativity (Pauli principle) and the quantum effects simply modify the relevant lengths squared. The cyclotron radius squared of the quasi-electron or the quasi-hole is modified by the amount~$\theta + 2C_{Q(e~or~h)}/(eB)$, which implies the minimum length squared of the cyclotron radius. In this way, both the noncommutativity and the quantum effects set the the minimum area in the quantum Hall effect. From the semi-classical discussions given in Sec. 1 and 2, we know for a simplest case,
\bea
\pi (r^2_{1d})_{min}= 2\pi (r^2_{2d})_{min}&=&2\pi \theta~~ (\mbox{noncommutativity or Pauli principle}), 
\label{98}
\\
\pi (r^2_{1d})_{min}= 2\pi (r^2_{2d})_{min}&=&\pi \frac{2}{eB} (2n+1)~~(\mbox{quantum effects}).
\label{99}
\eea

Our result in Eqs.(\ref{93}) and (\ref{94}) shows that these two kinds of effects of keeping the distance squared to be non-vanishing work also in our two component matrix model describing the quantum Hall effect. You can recognize that there is no $\theta$ term dominantly in the relative distance squared between quasi-electron and quasi-hole in Eqs.(\ref{93}) and (\ref{94}). This is mathematically achieved by $\lim_{N \rightarrow \infty} (\ln N)/(N+1)=0$, but is physically acceptable, since no Pauli principle works between electrons and holes.  Thus quantum effects work reasonably on the distance squared of the cyclotron radius of quasi-electron and of quasi-hole, as well as the distance squared between quasi-electron and quasi-hole. These consist of centrifugal force in the higher excitation modes and of the uncertainty from the zero point oscillations both of which are included in the number $C_{Q}$. in Eqs.(\ref{96}) and (\ref{97}).

Next, we derive the dispersion relation of the exciton from the above formula in Eq.(\ref{93}).  The dispersion relation is that of the energy in terms of the total momentum $P_{exc}$ of the exciton.  Therefore, we should express the cyclotron radii and the distance between quasi-electron and quasi-hole in terms of $P_{exc}$.  Classical picture of the exciton is the following: Lorentz forces causing the cyclotron motions of quasi-electron and quasi-hole are balanced by the Coulomb attraction, so that the three relevant vectors $\vec{x}_e-(\vec{x}_e)_{cm}$, $\vec{x}_h-(\vec{x}_h)_{cm}$, and $\vec{x}_e-\vec{x}_h$ are on the same line.  Here, $(\vec{x}_e)_{cm}$ and $(\vec{x}_h)_{cm}$ represent the center of the cyclotron motions. Then, both quasi-electron and quasi-hole move with the same velocities in the perpendicular direction to this line, forming a bound state of the exciton.

From the discussion in Sec. 2, the magnitude of the cyclotron momentum of the quasi-electron and quasi-hole $p_{qe}$ and $p_{qh}$ are given, respectively, by
\bea
p_{qe}&=&m_{qe}\frac{1}{\sqrt{2}}|z_e-(z_e)_{cm}|\omega_{qe}=\frac{1}{\sqrt{2}}(\nu eB)|z_e-(z_e)_{cm}|,
\label{100}
\\
p_{qh}&=&m_{qh}\frac{1}{\sqrt{2}}|z_h-(z_h)_{cm}|\omega_{qh}=\frac{1}{\sqrt{2}}(\nu eB)|z_h-(z_h)_{cm}|,
\label{101}
\eea
where the $\frac{1}{\sqrt{2}}$ is necessary to consider the motion in the "2d" picture.  
Force balancing is realized at the equilibrium configurations (classical solutions) of $E_{exc}$ in the classical approximation.  
These equilibrium configurations give
\bea
\frac{(\nu eB)^2}{2m_{qe}}|z_e-(z_e)_{cm}|
=\frac{(\nu e)^2}{4\pi\varepsilon}\frac{1}{|z_e-z_h|^2}
=\frac{(\nu eB)^2}{2m_{qh}}|z_h-(z_h)_{cm}|.
\label{102}
\eea
Therefore, in terms of the total momentum of the exciton $P_{exc}$, length squared can be expressed using Eqs.(\ref{100}),(\ref{101}) and (\ref{102}) as follows:
\bea
|z_e-(z_e)_{cm}|&=&\frac{\sqrt{2}m_{qe}}{\nu eBM}P_{exc},
\label{103}
\\
|z_h-(z_h)_{cm}|&=&\frac{\sqrt{2}m_{qh}}{\nu eBM}P_{exc},
\label{104}
\\
\frac{1}{|z_e-z_h|^2}&=&\frac{4\pi \varepsilon B}{\sqrt{2}\nu eM}P_{exc},
\label{105}
\eea
where $M=m_{qe}+m_{qh}$ is the total mass of the exciton.

Now, we obtain the dispersion relation of the exciton as follows:
\bea
E_{exc}&=& \frac{(P_{exc})^2}{2M}
+ \frac{(\nu eB)^2}{4m_{qe}}\left(\theta  + \frac{2}{eB} C_{Q(e)}\right) 
+ \frac{(\nu eB)^2}{4m_{qh}}\left(\theta  + \frac{2}{eB} C_{Q(h)} \right)
\nonumber \\
&-&
 \frac{(\nu e)^2\sqrt{\frac{4\pi \varepsilon B}{\nu eM}}}{4\pi \varepsilon}
  \sqrt{P_{exc}}
   \left( 1- \frac{1}{eB}\frac{4\pi \varepsilon B}{\nu eM} P_{exc} \left(C_{Q(e)} + C_{Q(h)}\right)+ \cdots\right),
\label{106}
\eea
where the Coulomb potential term can be also approximated with Eq.(\ref{94}) as follows:
\beq
-\frac{(\nu e)^2}{4\pi\varepsilon}
 \left(
  \frac{P_{exc}}
   {\frac{\nu eM}{4\pi \varepsilon B}
    +\frac{2}{eB}\left(C_{Q(e)}+C_{Q(h)}\right)P_{exc}}
 \right)^{\frac{1}{2}}
\label{107}
\eeq
This dispersion curve behaves $a-b\sqrt{P_{exc}}$ for small $P_{exc}$ with the certain constants $(a,b)$ and approaches to the classical dispersion relation $P_{exc}^2/2M$ for large $P_{exc}$. It takes a minimum point in between suggesting a roton excitation, or in other words the existence of a stable distance between the quasi-electron and quasi-hole, since the exciton energy can be written also as a function of the distance.
More detailed study is of course necessary to compare our results including the dispersion curve of the exciton to the experiments.
\section{Conclusion}
In this paper we propose a two component matrix model in the study of quantum Hall effect, in which two different infinite-sized matrices are prepared for the coordinates of electrons and holes, separately.  They evolve in time under the influence of the external magnetic field by the Lorentz force or the Chern-Simons type interactions.

We generalize the constraint condition of Susskind \cite{Susskind}, giving the Pauli principle for electrons and holes, in our model. We adopt in this paper a constraint under which the difference of areas occupied by electrons and holes preserves.  Accordingly, the surplus and the deficit of area can be realized separately, by using the electron matrix and the hole matrix. Then, the exciton solution of having both quasi-electron and quasi-hole is naturally obtained as a solution of the constraint condition. 

We study the quantization procedure of the matrix model naively by expanding the matrices around the classical solution of the exciton, and have given rather general physical states at the quantum level. They includes the Laughlin-like state as a special example. 

We estimate the energy of the exciton as a function of the coordinates of the locations of the quasi-electron and the quasi-hole, and afterwards the relation is rewritten as a function of the total momentum of the exciton, leading to the dispersion relation. In the calculation, there appear two kinds of corrections to the cyclotron radius squared of quasi-electron and quasi-hole, and to the distance squared between quasi-electron and quasi-hole.  One of the corrections comes from the noncommutativity of space or the Pauli principle, while the other one comes from the quantum effects. The latter corrections are related to the number of excited quanta in the background quantized state. The dispersion relation of the exciton obtained in this way gives a minimum, suggesting the possible magnetoroton mode or the existence of a stable distance between quasi-electron and quasi-hole. 

We have examined two different descriptions of the quantum Hall effect semi-classically, namely, the "1d" picture and the "2d" picture. Some of the correspondence between these two pictures are given. In this respect it is interesting to study the Calogero type "1d" description corresponding to our matrix model.  We will come back to this problem in the near future.

Our study of the quantum Hall effect using the two component matrix model is still at a primitive stage. More elaborate approximation methods developed for example in \cite{Exciton} and \cite{Kallin} should be incorporated, before doing the  numerical comparison of this work with the exciton experiments in the quantum Hall effect.

At the final stage of this work we have found a paper ~\cite{Coulomb} which studies the same topics as ours (the exciton problem), using the noncommutative Chern-Simons action. This is the other way to incorporate Pauli principle in the quantum Hall effect than our matrix model.  Relation between both methods is to be clarified.

\begin{center}
{\Large\bf Appendix:A\\ Difficulty of having exciton solution in 
the one matrix model}
\end{center}


 
In the introduction we have discussed the difficulty of having exciton
solution in the usual one matrix model.  To supplement the discussion there
we make a comment here on a possible "exciton solution" in the one matrix
model.  The fact that this "solution" violates the property of hermiticity
was a motivation of the authors to introduce the two matrix model in this
paper.
 
From the Susskind paper \cite{Susskind}, we know that the solution in Eqs.(\ref{sec2.15})
 gives a
quasi-particle having the surplus of the area,
\begin{equation}
[\hat{b}, \hat{b}^{\dagger}]=1 + \nu |0><0|.
\end{equation}
 
To add a quasi-hole and obtain an exciton solution, we need to generate the
deficit of area for the quasi-hole, in addition to the surplus of area for
the quasi-particle.  For this purpose, we span the infinitedimensional
vector space by $\{|-\infty>, \cdots, |-1>,|0>, |1>, \cdots, |\infty>\}$,
and assume the operators $\hat{b}$ and $\hat{b}^{\dagger}$ to satisfy the
usual relation [ Eqs.(\ref{sec2.15}) ]  in Section 2 for the quasi-particle, as well as
the following new relation for the quasi-hole:
\bea
\hat{b}^{\dagger}|-1>&=&0,  \nonumber
\\
\hat{b}^{\dagger}|-1-n>&=& i\sqrt{n-\nu}|-n> ~~\mbox{for}~~ n \ge 1,
\nonumber
  \\
\hat{b}|-1-n>&=& i\sqrt{n+1-\nu}|-2-n> ~~\mbox{for}~~ n \ge 0.
\label{appendix.2}
\eea
Then, we have a "solution" , satisfying
\begin{equation}
[\hat{b}, \hat{b}^{\dagger}]=1 + \nu |0><0|-\nu |-1><-1|,
\end{equation}
giving both the surplus and the deficit of area, so that this solution
seems to be a correct exciton solution.

However, we can immediately recognize that the hermitian conjugation
property between $\hat{b}$ and $\hat{b}^{\dagger}$ is violated for the hole
part by the existence of the extra "i" in Eq. (\ref{appendix.2}).   That is, the
following relations which should be hermitian conjugate with each other,
$\hat{z} \equiv \hat{x}^{1}+ i \hat{x}^{2}=\sqrt{2\theta}\hat{b}$ and
$\hat{z}^{\dagger} \equiv  \hat{x}^{1}- i
\hat{x}^{2}=\sqrt{2\theta}\hat{b}^{\dagger}$, become inconsistent.
 
Here, we can find that the state $|n>$ with negative $n$ can be considered
as the negative "energy levels below the Dirac sea", and in such a case the
creation and annihilation operators for the negative "energy levels"should
be replaced by the annihilation and creation operators of the
anti-particles, respectively.
Hence, we introduce the operators of the anti-particles (holes), $\hat{d}$
and $\hat{d}^{\dagger}$, defined by
\begin{eqnarray}
  \hat{d}&=& -i \hat{b}^{\dagger} \nonumber, \\
\hat{d}^{\dagger}&=& -i \hat{b},
\end{eqnarray}
and discard the negative "energy levels", by replacing them by the positive
"energy levels" as $|-1-n> \to |n>~~(n \ge0) $.  Then the relation in Eq.
(\ref{appendix.2}) is transformed to the relation [Eqs. (\ref{sec2.16})]
 given in Section 2.
 
At this stage the wrong hermiticity for the hole part is properly
recovered,and we arrive at the two matrix models of Section 2.  Namely,
corresponding to the different sets of operators $(\hat{b},
\hat{b}^{\dagger})$ and $(\hat{d}^{\dagger}, \hat{d})$, we have to
preparefrom the beginning, the different matrices, $\hat{x}_{e}$ and
$\hat{x}_{h}$ for electron (particle) and hole (anti-particle),
respectively.  In this prescription, the degrees of freedom is not doubled,
since what we have done is only to use the different operators to express
the negative "energy levels".

%
%
\section*{Acknowledgment}
The authors are grateful to Kenzo Ishikawa for his series of lectures on quantum Hall effect, and Bunji Sakita for his seminar at Ochanomizu U..  We also give our thanks to Hiromi Yakazu for her initial collaboration.  
One of the author (A.S.) is supported in part by the Grant-in-Aid for Scientific Research from the Ministry of Education and Science (No.11640262).  The other one (B.P.M.) is supported by the JSPS postdoctral fellowship (No. P01022).

\end{document}